\documentstyle[12pt,aasms4,epsf]{article}
\def\be{\begin{equation}}
\def\ee{\end{equation}}

\def\today{\ifcase\month\or January\or February\or March\or April\or
May\or
June\or July\or August\or September\or October\or November\or
December\fi
  \space\number\day, \number\year}

\def\ARAA{Ann. Rev. Astron. Astrophys.}

\def\etal{{\it et al.\ }}

\def\cf{{cf.\ }}
\def\eg{{e.g.}}
\def\ie{{i.e.}}

\def\chisq{$\chi^2$}
\def\LCDM{$\Lambda$-CDM}
\def\TCDM{T-CDM}

\def\ltsima{$\; \buildrel < \over \sim \;$}
\def\lsim{\lower.5ex\hbox{\ltsima}}
\def\gtsima{$\; \buildrel > \over \sim \;$}
\def\gsim{\lower.5ex\hbox{\gtsima}}
\def\ga{\mathrel{\hbox{\rlap{\hbox{\lower4pt\hbox{$\sim$}}}\hbox{$>$}}}}
\def\la{\mathrel{\hbox{\rlap{\hbox{\lower4pt\hbox{$\sim$}}}\hbox{$<$}}}}

\def\ifm#1{\relax\ifmmode#1\else$\mathsurround=0pt #1$\fi}

\def\la{\langle}

\def\pmb#1{\setbox0=\hbox{#1}%
 \kern-.025em\copy0\kern-\wd0
 \kern.05em\copy0\kern-\wd0
 \kern-.025em\raise.0433em\box0}

\def \r0p{ r{_0^\prime}}

\def \m3{{\rm Mark III}}

\begin{document}

\title{Goodness-of-Fit Analysis of Radial Velocities Surveys}
\author{Y. Hoffman  \altaffilmark{1}  and S. Zaroubi \altaffilmark{2}}
\altaffiltext{1}{Racah Institute of Physics, The Hebrew University,
Jerusalem 91904, Israel}
\altaffiltext{2}{Max Planck Institute for Astrophysics,
Karl-Schwarzschild-Str. 1, D-85740 Garching, Germany}

\begin{abstract}
Using eigenmode expansion of the \m3\ and SFI surveys of cosmological
radial velocities a goodness-of-fit analysis is applied on a
mode-by-mode basis. This differential analysis complements the
Bayesian maximum likelihood analysis that finds the most probable
model given the data. Analyzing the surveys with their corresponding
most likely models from the CMB-like family of models, as well as with
the currently popular \LCDM\ model, reveals a systematic inconsistency
of the data with these `best' models. There is a systematic trend of
the cumulative \chisq\ to increase with the mode number (where the
modes are sorted by decreasing order of the eigenvalues). This
corresponds to a decrease of the \chisq\ with the variance associated
with a mode, and hence with its effective scale. It follows that the
differential analysis finds that on small (large) scales the global
analysis of all the modes `puts' less (more) power than actually
required by the data.  This observed trend might indicate one of the
followings: a.  The theoretical model (\ie\ power spectrum) or the
error model (or both) have an excess of power on large scales;
b. Velocity bias; c. The velocity data suffers from still uncorrected
systematic errors.

\end{abstract}


\section{Introduction}

Surveys of radial velocities of galaxies have played a major role in 
the study of the large scale structure.  The analysis of such surveys 
has been conducted in two main directions, the mapping of the local 
cosmography and the estimation of the cosmological parameters (\cf\ 
Dekel 1994 for a review).  The Bayesian framework provides one with 
very elegant and powerful tools for conducting both the mapping and 
parameter estimation, where the recovery of the large scale structure 
is done by means of the Wiener filter and the parameters are estimated 
by maximum likelihood (MaxLike) analysis (Zaroubi \etal\ 1995, 
hereafter ZHFL).  
In the case where the deviations from a homogeneous and isotropic 
universe constitute a Gaussian random field the Wiener filter and the 
MaxLike are the optimal tools for performing such an analysis (ZHFL).  
Indeed, the MARK III catalog of radial velocities (Willick \etal\ 
1995, 1996, 1997a) have been recently analyzed by Wiener filtering 
(Zaroubi, Hoffman and Dekel 1999) and by MaxLike (Zaroubi \etal\ 
1997).  The SFI survey of da Costa \etal\ (1996) has been studied by 
MaxLike analysis by Freudling \etal\ (1999) and by Wiener filtering 
(Hoffman and Zaroubi, unpublished).  Both surveys seem to yield 
similar results.

In the Bayesian MaxLike analysis one calculates the posterior 
probability of a model to be correct given the data (ZHFL, Vogeley and 
Szalay 1996).  Thus the model that maximizes the likelihood function, 
over a given parameter (or model) space, is the most likely model in 
that space.  The MaxLike analysis cannot guarantee, however, that the 
most probable model is indeed consistent with the data.  It provides 
only a relative measure for models to be correct.  It is common to 
adopt an independent measure for the goodness-of-fit, which is often 
given by the requirement that the reduced $\chi^{2}$  is close to 
unity.  Often, when the most likely model (given the data) passes also 
the goodness-of-fit test one assumes that the `correct' model has been 
nailed down.  Here, the $\chi^{2}$ test is expanded and a much more 
critical test is suggested and then applied to the \m3\ and SFI 
surveys.

The \chisq\ 'goodness-of-fit' is based on the assumptions that all the 
random variables that affect the observables are normally distributed.  
In the cosmological context this applies to both the underlying 
dynamical (\eg\ density and velocity) field and the statistical 
errors.  Thus for a survey containing $N$ data observables (\eg\ 
radial velocities) the \chisq\ of the system of $N$ degrees of freedom 
(DOF) is calculated given the model that maximizes the likelihood 
function.  The goodness-of-fit is measured by how close is the 
$\chi^{2}/{\rm DOF}$ to unity.  This provides a global measure for the 
consistency of the data with the model, as it includes all the 
observables.  A situation might occur of some `conspiracy' where 
different parts of the data   deviate from the predictions of the 
model, but when combined together they `conspire' to yield a 
reasonable \chisq.  A much stronger test on the model is to decompose 
the data into statistical independent eigenmodes and observe the 
\chisq\ behavior of the independent modes.  Eigenmode analysis, also 
known as principal component analysis (PCA) and the Karhunen-Loeve 
transform, is not a new tool in the field.  It has been applied to 
studies of redshift surveys (Vogeley and Szalay 1996),the cosmic 
microwave background (Bunn 1997, Bond 1995) and more recently radial 
velocities surveys (Hoffman, 1999).  The later study is extended here 
to perform the 'goodness-of-fit' test on a mode-by-mode basis.  The 
basic formalism is presented in \S~\ref{sec:method}, and its 
application to the \m3\ and SFI surveys is given in 
\S~\ref{sec:chisq}.  Our results are discussed and the conclusions are 
summarized in \S~\ref{sec:discussion}.

\section{Eigenmode Analysis of Radial Velocities }
\label{sec:method}

%Radial velocities

Consider a data base of radial velocities $\{ u_{i}\}_{i=1,\ldots,N}$, 
where 
\begin{equation}
u_{i} = {\bf  v}
({\bf r}_{i})  \cdot \hat {\bf r}_{i} 
+\epsilon_{i},
	\label{eq:ui}
\end{equation}
${\bf  v}
$ is the three dimensional velocity, ${\bf r}
_{i}$ is the 
position of the i-th data point  and $\epsilon_{i}$ is the statistical 
error associated with the i-th radial velocity. The assumption made 
here is of   a   cosmological model that  well 
describes the data, that systematic errors have been properly dealt 
with and that the statistical errors are well understood. 
The data auto-covariance matrix  is then  written as:
\begin{equation}
R_{ij} \equiv
\Bigl < u_i u_j  \Bigr > =   \hat {\bf r}_j \Bigl < {\bf  v}
({\bf r}_i) {\bf  v} ({\bf r}_j)  \Bigr > 
   \hat {\bf r}_j   + \sigma{^2_{ij}}.
\label{eq:Rij}
\end{equation}
(Here $\Bigl <  \ldots  \Bigr >$ denotes an ensemble average.)
The last term is the error covariance matrix.
The velocity covariance tensor that enters this equation was derived 
by  G\'orski (1988, see also Zaroubi, Hoffman and Dekel 
1999) and it depends on the power spectrum and cosmological 
parameters.

The eigenmodes of the data covariance matrix provides a natural 
representation of the data:
\begin{equation}
R {\bf\eta}^{(i)}  =\ \lambda_{i} {\bf\eta}^{(i)}
\label{eq:eigenvec}
\end{equation}
The set of $N$ eigenmodes   $\{ {\bf \eta}^{(i)}\}$ constitutes an 
orthonormal basis and the eigenvalues $\lambda_{i}$  are arranged in 
decreasing order (in absolute values). A new representation of the data   
is given by:
\begin{equation}
\tilde a_{i} = \eta{^{(i)}_{j}} \  u_{j}
\label{eq:aiui}
\end{equation}
This provides a 
statistical orthogonal representation, namely:
\begin{equation}
\bigl\langle \tilde a_{i} \tilde a_{j} \bigr\rangle \ = 
 \ \lambda_{i} \delta_{ij} 
\label{eq:tildeaiaj}
\end{equation}
The normalized transformed variables are defined by:
\begin{equation}
a_{i}  = {\tilde a_{i}  \over \sqrt{\lambda_{i}} }
\label{eq:defai}
\end{equation}
Eq. \ref {eq:tildeaiaj} is written now as:
\begin{equation}
\bigl\langle   a_{i}   a_{j} \bigr\rangle \ = 
 \ \delta_{ij} 
\label{eq:aiaj}
\end{equation}
Note that as the modes are statistically independent one can measure 
the $\chi^{2}$ of a given mode,
%\begin{equation}
$\chi{^{2}_{i}}=a{^{2}_{i}}$,
%\label{eq:chi2i}
%\end{equation}
and the cumulative reduced \chisq\ is given by:, 
\begin{equation}
\chi{^{2}_{M}}={1\over M}\sum_{i=1}^{M}a{^{2}_{i}}
\label{eq:chi2m}
\end{equation}
 For normally distributed errors and a Gaussian random velocity field 
the $a_{i}$'s are normally distributed with zero mean and a variance 
of unity.

 In addition the probability 
of finding such $\chi{^{2}_{M}}$ is calculated  as well. The probability 
is defined by
\begin{eqnarray}
P\Bigl( \chi{^{2}_{M}} \Bigr) & = &\; \; P_{\chi^{2}}(M 
\chi{^{2}_{M}},M) \;\;\;\;\;\; \;  {\rm for} \; P_{\chi^{2}}(M 
\chi{^{2}_{M}},M) < 0.5 \cr
          & = &\; \; 1 -  P_{\chi^{2}}(M \chi{^{2}_{M}},M)
               \;\;\; \; \; \;  {\rm otherwise},   
\label{eq:chi2i}
\end{eqnarray}
where $P_{\chi^{2}}( x,M)$ is the   probability  that a random 
variable drawn from a  \chisq\  
distribution with $M$ degrees of freedom   is less than  a given value $x$.

%%%%%%%%%%%%%%%%%%%%%%%%%%%%%%
\section{Differential \chisq\  Analysis}
\label{sec:chisq}

Here the  goodness-of-fit  of the \m3\ and SFI surveys is studied. The 
models studied here are the MaxLike solutions for these surveys, which 
are slightly different from one another. The most likely model given 
\m3\ is a tilted-CDM (\TCDM) of $\Omega_{0}€=1,\ h=0.75$ and $n=0.8$ 
where $\Omega_{0}$ is the cosmological density parameter, $h$ is 
Hubble's constant in units of $100\ km/s/Mpc$ and $n$ is the power 
spectrum index  (Zaroubi \etal\ 1977). The most likely model given SFI 
is an open CDM (OCDM) of  $\Omega_{0}€=0.79,\ h=0.6$ and $n=0.92$ 
(Fruedling \etal 1999). For both cases the MaxLike best model has a 
total $\chi{^{2}_{M=N}}$ very close to unity. Thus, from the point of 
view of the integral \chisq\ the MaxLike solutions seem to be very 
consistent with the data. This is extended to perform a 
differential \chisq\ analysis, namely to study the \chisq\ behavior 
across the modes spectrum.

To study the robustness of this probe it is first applied  to a 
linear mock catalog of \m3, constructed from an unconstrained 
realization of the   velocity field.  This field is sampled 
at the location of the \m3\ data points, to which normally 
distributed errors are added according to \m3's error covariance matrix.  
The cumulative \chisq\ of such a catalog should oscillate around unity, 
given that the model used to generate the catalog is known.  Indeed, 
this has been confirmed by an analysis of a few linear mock catalogs of 
\m3. The probabilities of obtaining such \chisq\ distribution lies 
comfortably within the $90\% $ confidence
level.  The non-trivial result of this test is that the very poor 
sampling of the long wavelength Fourier waves, \ie\ cosmic variance, 
does not affect the goodness-of-fit  test.

\begin{figure} 
\plottwo{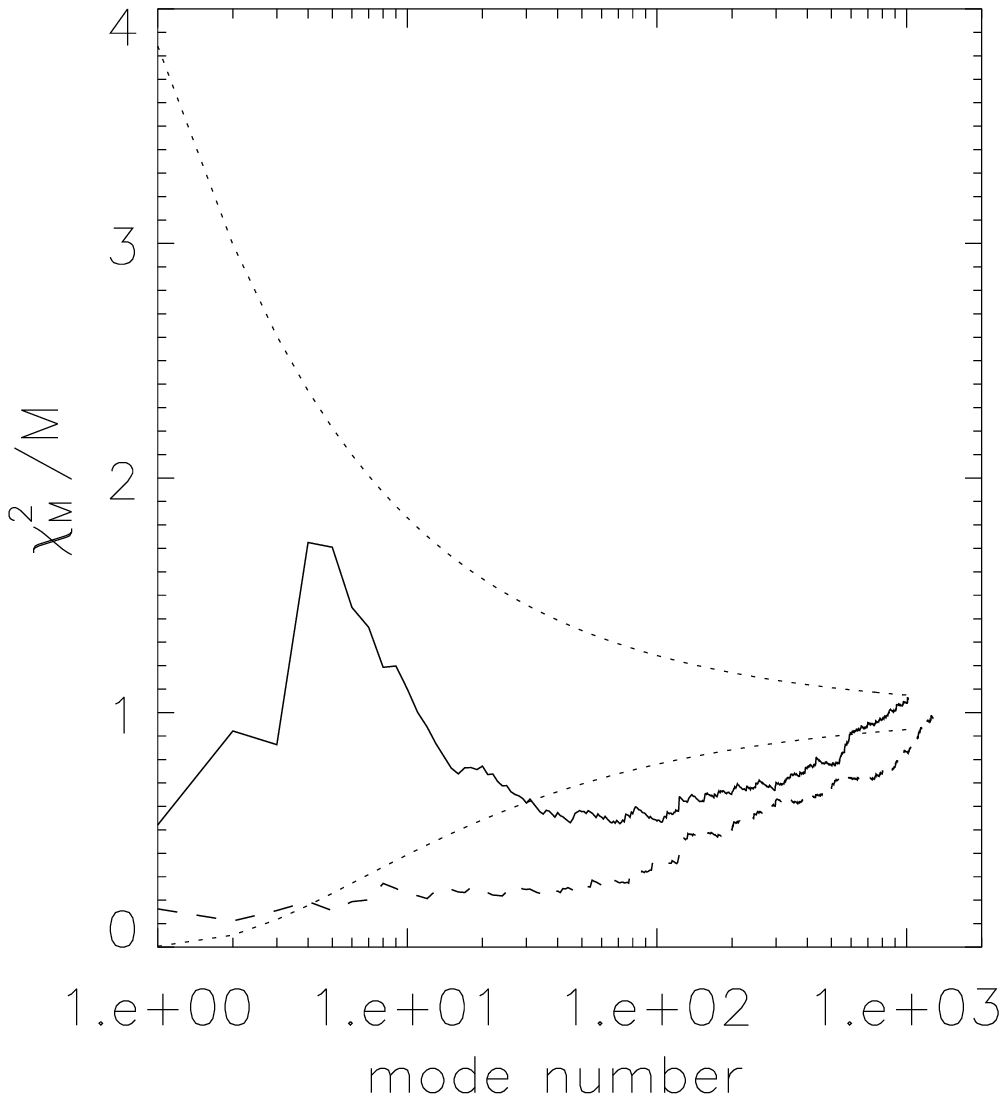}{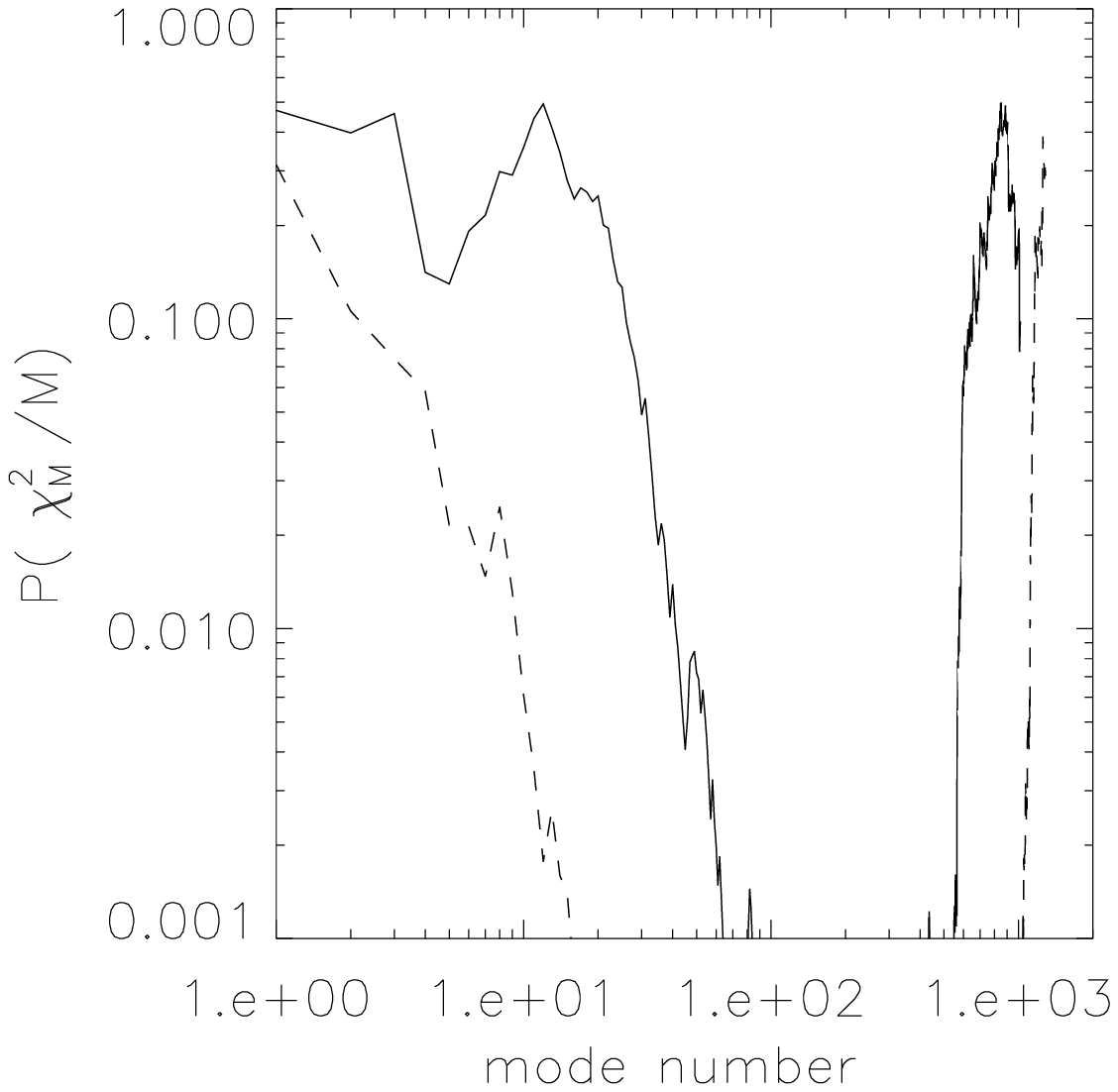}
\caption{The cumulative \chisq\ (left) and 
the probability of this \chisq\ distribution (right) of the \m3\ (solid 
line) and 
SFI (dashed line)  surveys are plotted 
against the mode number. 
The model used here are the tilted-CDM 
($\Omega_{0}€=1,\ h=0.75$ and $n=0.8$; \m3) 
and the open CDM   ($\Omega_{0}€=0.79,\ h=0.6$ and $n=0.92$; SFI)
The lower and upper 90\% confidence levels are superimposed on the 
left figure (dotted lines). The modes are arranged by decreasing order.} 
\label{fig:fig1_m3_sfi_ml}
\end{figure}

The differential \chisq and its associated probability of the \m3\ and 
SFI surveys are presented in Fig. \ref{fig:fig1_m3_sfi_ml}, each case 
analyzed in its maximum likelihood solution. A clear trend is noticed, 
namely over almost the entire mode spectrum the cumulative \chisq\ 
increases monotonically. When all modes are included the   total 
\chisq/DOF is indeed close to $1$, but if we had to take half the 
modes, starting from the top or the bottom, a very different \chisq\ 
would have obtained. 

%M3 - TCDM, SFI OCDM

The differential \chisq\  analysis is repeated for the currently 
popular model of \LCDM ($\Omega_{0}=0.4,\ h=0.6$ and $n=1$; 
Fig. \ref{fig:fig1_m3_sfi_ml}). 
Indeed, the same trend is found in this case as well 
but the total \chisq converges to a value outside the $90\% $ 
confidence level.

%M3 SFI LCDM
\begin{figure} 
\plottwo{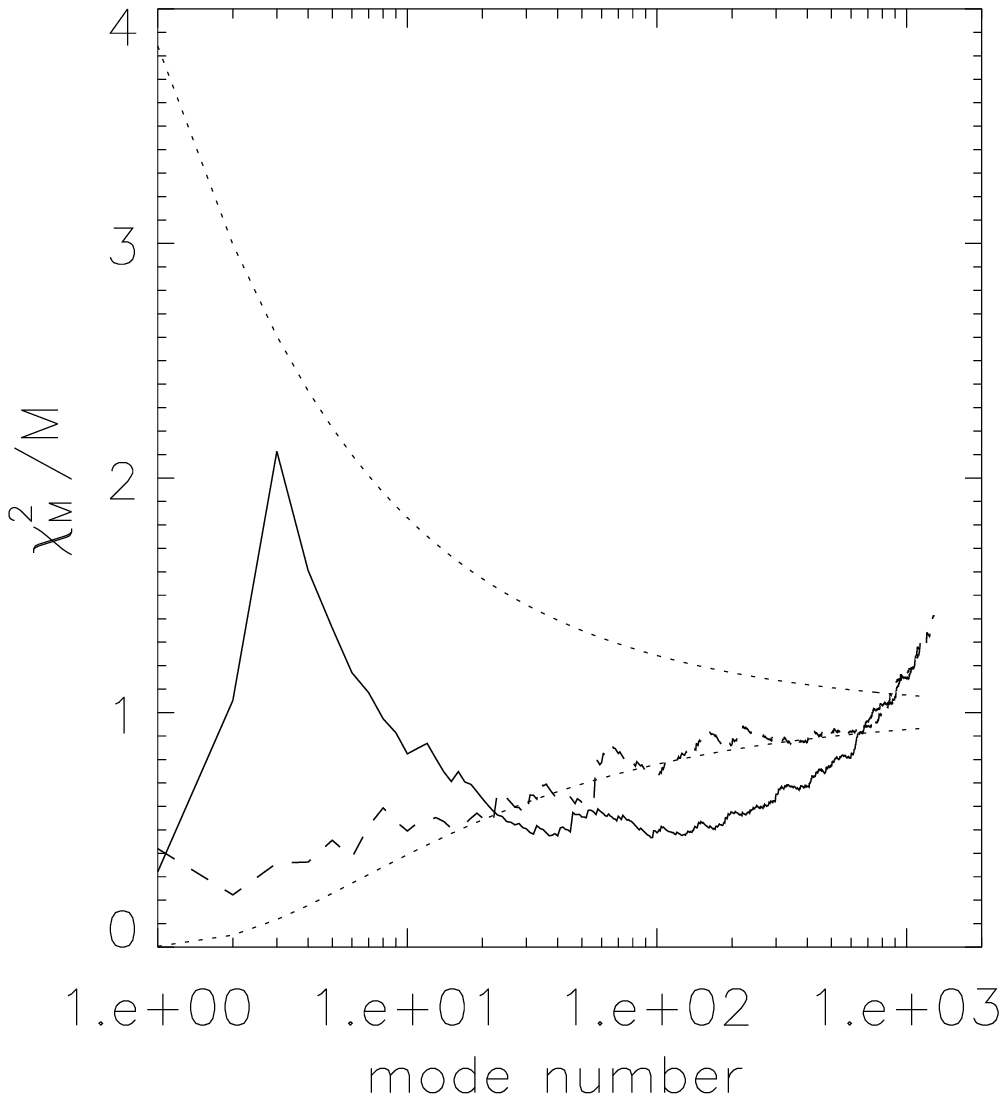}{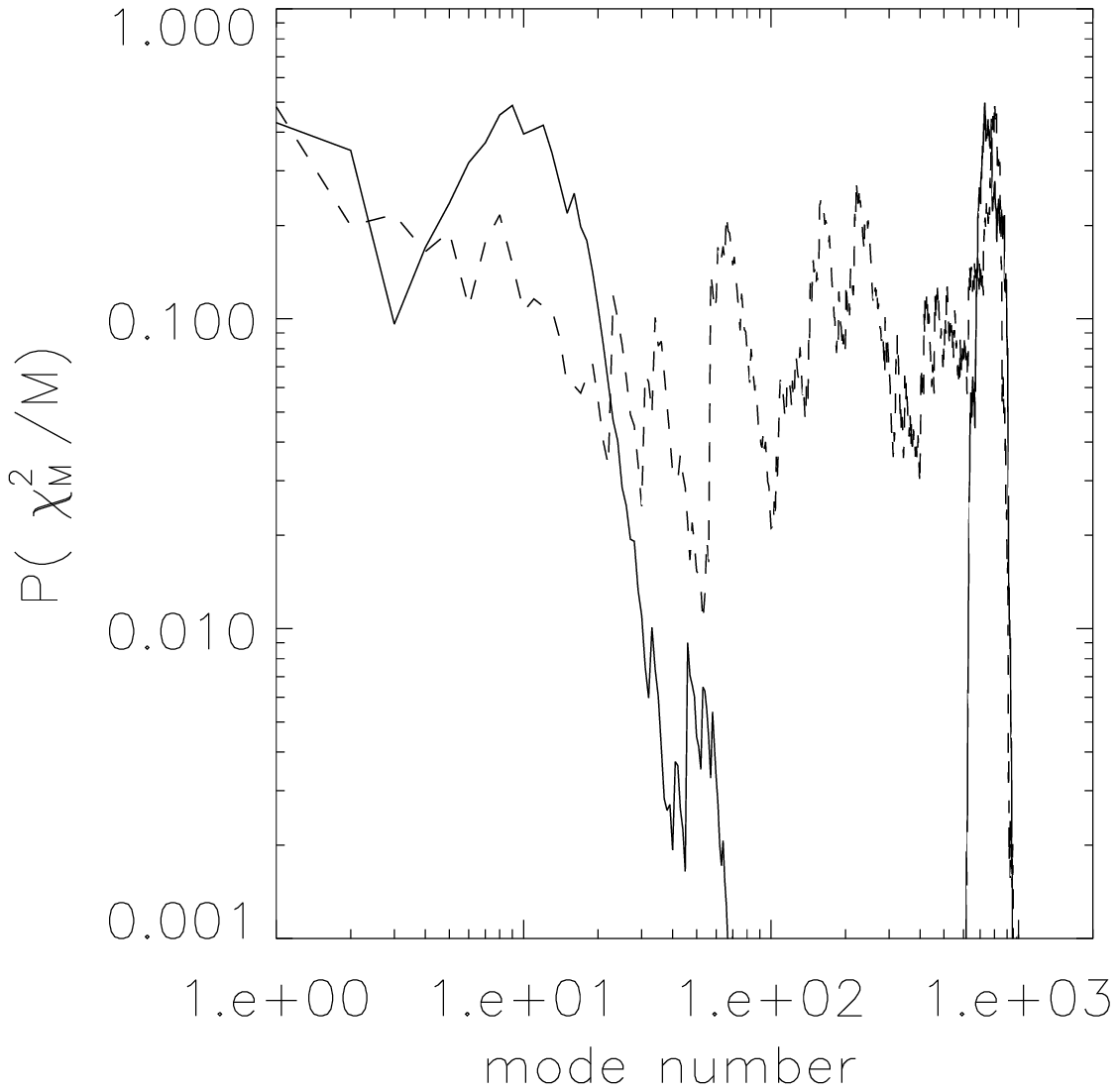}
\caption{Same analysis as in \ref{fig:fig2_m3_sfi_ml} 
applied to the \m3\ and SFI survey. The model is a \LCDM
 ($\Omega_{0}€=0.4,\ h=0.6$ and $n=1$)}
\label{fig:fig2_m3_sfi_lcdm}
\end{figure}

The conclusions that follows is that  for both data sets, \m3\ 
and SFI, and for a variety of theoretical models the differential  
\chisq\ increase monotonically with the mode number (with  the 
exception of the first 10 modes of the \m3). The theoretical 
expectation is that if indeed the data is consistent with the assumed 
model then   $\chi{^{2}_{M}}$ will fluctuate around unity. The 
probability of observing such a trend  given a model is very small 
across most of the mode number range.

\section{Discussion}
\label{sec:discussion}

%Inconsistency of data with best CDM-like models: Theoretical model + 
% error model 

What have we learned from the differential \chisq\ analysis?  
It has been found that even the most probable CDM-like model, the one 
that maximizes the likelihood function given the data, is not fully 
consistent with the data.  The cumulative \chisq\ has been calculated 
both downwards and upwards (namely starting from the modes with the 
largest and smallest eigenmode, respectively).  Over more than $90\% $ 
of the modes the cumulative \chisq\ lies well outside the $90\% $
confidence level, indicating a very small probability of measuring 
such data given the assumed model.  Over most of the mode number range 
$\chi{^2_M}$ increases monotonically.  It is this   behavior 
of the \chisq\ which indicates a systematic inconsistency of the model 
with the data.  The assumed model actually contains two ingredient, 
the theoretical power spectrum and the error model.  However, the 
present analysis cannot indicate which one is to be `blamed' for the 
systematic trend.  It should be noted here that apart from the first 
few ($10 - 20$) modes there is a clear correlation of the eigenvalues 
with its weighted mean distance (of data points of the given mode).  
Namely, the variance associated with a mode (\ie\ its eigenvalue) 
increases with its mean distance (Zehavi, private communication, 
Silverman \etal\ in preparation).  It follows that the \chisq\ trend 
seen here is closely correlated with the distance and that the data 
`asks' for less power on large scales than the model (power spectrum 
and noise) provides.  A detailed study of the power spectrum and error 
model possible modifications is to be given elsewhere (Silverman 
\etal\ in preparation).  (Note that these first $10 - 20$ modes are 
the ones dominated by the underlying velocity field and not the noise, 
Hoffman 1999.)

The cosmological implications of the present findings are that either 
the error  and/or the theoretical model need to be 
modified.  The theoretical model assumed in the analysis of large 
scale radial velocity surveys is that the velocities are drawn from a 
Gaussian random field defined by a given power spectrum. The present 
study might indicate the inconsistency of the power spectrum with the 
data. A less likely possibility is that it indicates  a departure from 
the Gaussian statistics. Alternatively, the present work might indicate a 
systematic error that has not been accounted for that causes this 
trend. Still another possibility is that of an indication for a 
velocity bias. 

The conclusions reached here should not be taken as a contradiction of 
the results of Zaroubi \etal\ (1997) and Fruedling \etal\ (1999), but rather 
as  extending and complementing them.  The Bayesian MaxLike analysis can be performed 
only within the assumed parameter/model space.  The differential 
\chisq\ allows one to go beyond this and analyze the nature of the 
agreement, or the lack of it, between a given model and the data on a 
mode by mode basis.

%Methodology

 The PCA transforms the data to a statistically independent 
 representation and enables the study of the compatibility of the data 
 with the model on a mode by mode basis.  This differential analysis 
 is in contrast to the more traditional approach where a  data 
 set is analyzed as a whole.  The differential \chisq\ analysis should 
 be performed together with the Bayesian MaxLike analyze and 
 complement it.  This should be useful in fields where the MaxLike is 
 the basic tool of analysis such as the mapping of the CMB angular 
 fluctuations and the study of redshift surveys as well as all radial 
 velocity surveys.  The present analysis can prove to be very useful 
 and powerful in those fields where systematic errors play a crucial 
 roles, such as redshift and radial velocities surveys.

\acknowledgments We have benefited from many interesting discussions 
with Avishai Dekel, Zafrir Kolatt, Ofer Lahav, Lior Silverman, Simon 
White and Idit Zehavi.  The hospitality of the Racah Inst.  Physics 
and the Max Planck Institut fur Astrophysik is gratefully 
acknowledged.  This research has been partially supported by a 
Binational Science Foundation grant 94-00185 and an Israel Science 
Foundation grant 103/98.

{}

\clearpage

\end{document}